\documentclass[aps,pre,superscriptaddress]{revtex4}
\usepackage{epsfig}
\usepackage{times}
\usepackage{amsmath}
\usepackage{amssymb}
\usepackage{float}
\begin{document}

\title{Is type 1 diabetes a chaotic phenomenon?}

\author{Jean-Marc Ginoux}
\email{ginoux@univ-tln.fr}
\affiliation{Laboratoire d'Informatique et des Syst\`{e}mes, UMR CNRS 7020, CS 60584, 83041 Toulon Cedex 9, France}

\author{Heikki Ruskeep\"{a}\"{a}}
\email{ruskeepaa@utu.fi}
\affiliation{University of Turku, Department of Mathematics and Statistics, FIN-20014 Turku, Finland}

\author{Matja{\v z} Perc}
\email{matjaz.perc@uni-mb.si}
\affiliation{Faculty of Natural Sciences and Mathematics, University of Maribor, Koro{\v s}ka cesta 160, SI-2000 Maribor, Slovenia}
\affiliation{CAMTP -- Center for Applied Mathematics and Theoretical Physics, University of Maribor, Mladinska 3, SI-2000 Maribor, Slovenia}
\affiliation{Complexity Science Hub, Josefst{\"a}dterstra{\ss}e 39, A-1080 Vienna, Austria}

\author{Roomila Naeck}
\affiliation{PSASS, Ecoparc de Sologne, Domaine de Villemorant, 41210 Neug-sur Beuvron, France}

\author{V\'{e}ronique Di Costanzo}
\affiliation{Centre Hospitalier Intercommunal de Toulon La Seyne, 54, rue Henri Sainte Claire Deville, CS31412, 83056 Toulon Cedex, France}

\author{Moez Bouchouicha}
\affiliation{Laboratoire d'Informatique et des Syst\`{e}mes, UMR CNRS 7020, CS 60584, 83041 Toulon Cedex 9, France}

\author{Farhat Fnaiech}
\affiliation{Universit\'{e} de Tunis, Laboratory of Signal Image and Energy Mastery, ENSIT, Avenue Taha Hussein 1008 Montfleury, Tunisia}

\author{Mounir Sayadi}
\affiliation{Universit\'{e} de Tunis, Laboratory of Signal Image and Energy Mastery, ENSIT, Avenue Taha Hussein 1008 Montfleury, Tunisia}

\author{Takoua Hamdi}
\affiliation{Universit\'{e} de Tunis, Laboratory of Signal Image and Energy Mastery, ENSIT, Avenue Taha Hussein 1008 Montfleury, Tunisia}

\begin{abstract}
A database of ten type 1 diabetes patients wearing a continuous glucose monitoring device has enabled to record their blood glucose continuous variations every minute all day long during fourteen consecutive days. These recordings represent, for each patient, a time series consisting of 1 value of glycaemia per minute during 24 hours and 14 days, i.e., 20,160 data point. Thus, while using numerical methods, these time series have been anonymously analyzed. Nevertheless, because of the stochastic inputs induced by daily activities of any human being, it has not been possible to discriminate chaos from noise. So, we have decided to keep only the 14 nights of these ten patients. Then, the determination of the time delay and embedding dimension according to the delay coordinate embedding method has allowed us to estimate for each patient the correlation dimension and the maximal Lyapunov exponent. This has led us to show that type 1 diabetes could indeed be a chaotic phenomenon. Once this result has been confirmed by the determinism test, we have computed the Lyapunov time and found that the limit of predictability of this phenomenon is nearly equal to half the 90-minutes sleep-dream cycle. We hope that our results will prove to be useful to characterize and predict blood glucose variations.
\end{abstract}

\maketitle

\section{Introduction}
\noindent

Diabetes is a chronic disease which affects more than two hundred millions of people in the world \cite{Wilde2004}. There are two main types of diabetes: type 1 and type 2. In the case of type 1 diabetes, the lack of insulin due to the destruction of insulin-producing beta cells in the pancreas leads to diabetes mellitus in which ``insulin is required for survival'' to prevent the development of ketoacidosis, coma and death. Type 2 diabetes is characterized by disorders of insulin action and insulin secretion, including insulin resistance \cite{Alberti1998}. This work only concerns type 1 diabetes. Insulin-dependent diabetes or type 1 diabetes is characterized by dramatic and recurrent variations in blood glucose levels. The effects of such variations are irregular (erratic) and unpredictable hyperglycemias, sometimes involving ketoacidosis, and sometimes serious hypoglycemias. Nowadays, the development of Continuous Glucose Monitoring (CGM) devices makes it possible to record blood glucose every minute during two weeks providing endocrinologists thousands of data in the form of time series. This has also led to prediction of continuous blood glucose variations based on computational methods such as support vector machine \cite{Hamdi2018}.

In the middle of the eighties, Wolf et al. [1985] proposed an algorithm allowing the ``estimation of non-negative Lyapunov exponents from an experimental time series''. Thus, determination of Lyapunov exponents enabled, on the one hand, to decide whether the time series is chaotic or not and, on the other hand, to assess the Lyapunov time corresponding to the limit of predictability of the observed phenomenon. Since this algorithm was first published in Fortran code, many other versions have been developed in various languages such as C and C++ \cite{Kodba2005}. Recently, this Fortran code has been implemented in MatLab by Wolf, as well as in Mathematica by Ruskeep\"{a}\"{a} in 2014, see \url{http://library.wolfram.com!infoce nter/MathSource/8775/}. This Mathematica software is used in the present work. The software is shortly explained in [Ruskeep\"{a}\"{a}, 2015] and is also used in \url{http://demonstrations.wolf'ram.com/ChaoticDataMaXimaILYapunovExponent/} (2017).

Long after the famous French mathematician, physicist and engineer, Henri Poincar\'{e} had discovered ``deterministic chaos'' in his seminal works concerning the motion of celestial bodies \cite{Poin1892}, many scientists searched for traces of chaotic behavior in various phenomena. The signature of chaos is the very well-known property of ``Sensitive Dependence upon Initial Conditions'' (SDIC) which makes the observed phenomenon unpredictable in long term. In the beginning of the sixties, Edward Norton Lorenz was the very first to identify such a feature in Meteorology \cite{Lorenz1963}. More than ten years later, Sir Robert May demonstrated the existence of ``complex dynamics'' (chaos) in ecological models \cite{May1976}. During the following decades, scientists highlighted chaotic behavior in human body. Indeed, according to Ives [2004] ``innumerable, entwined (nonlinear) feedback loops regulate our internal processes, keeping us within the narrow bounds needed for survival. Despite this regulation, our systems are aperiodic and unpredictable in the long term.'' Thus, a prime example of chaos was found in the brain \cite{bablo85} and then, in the beating of the heart \cite{dibern98, sign98}. Although type 1 diabetes is widely and intuitively considered by endocrinologists and clinicians as a chaotic phenomenon \cite{Goldstein1997, Berg2008, Little2011}, this has not yet been established by numerical methods to our knowledge.

In this work, starting from a database of glucose from ten type 1 diabetes patients and while using well-known numerical algorithms with Mathematica, we give support to the conclusion that type 1 diabetes is a chaotic phenomenon and we provide the Lyapunov time corresponding to the limit of predictability of this phenomenon. These results will prove to be very useful to characterize and predict blood glucose variations. "-
This paper is organized as follows. In Section 2.1, we briefly present the three main continuous glucose monitoring devices and detail the main features of the one used in this study. In Section 2.2, we recall the method of time delay reconstruction, also referred to as delay reconstruction and phase space reconstruction \cite{packard1980, takens1981}, and the definitions of ``time delay'' and ``embedding dimension''. To determine their proper values we use, respectively, ``average mutual information'' \cite{fraser1986} and the method of ``false nearest neighbors'' proposed by Kennel et al. \cite{kennel1992}. We recall the method of correlation exponent \cite{Grassberger1983} to estimate the correlation dimension and the method of local divergence rates \cite{kantz1994} to estimate the maximal Lyapunov exponent. In Section 3.1, we present, in some detail, all the results for the glucose data of patient 1. Then, in Section 3.2, we briefly summarize the results of all the ten patients. We also apply a direct test for determinism in a time series \cite{kaplan1992} and the programs by Perc to state that type 1 diabetes is likely a chaotic phenomenon.
Section 4 summarizes the results of the article {\it i.e.}, that an estimation of the Lyapunov time which is nearly equal to half the 90-minutes sleep-dream cycle.

\section{Materials and Methods}

\subsection{Continuous Glucose Monitoring Systems}

Continuous Glucose Monitoring (CGM) devices began to be developed in the eighties. However, they became available for practical use only twenty years ago with the miniaturization and development of electronic sensors combined to growing storage capacities. These systems, which have been proven to reliably reflect glucose levels \cite{bailey2015}, replace henceforth the classical finger prick blood glucose readings by monitoring interstitial fluid (ISF) glucose levels continuously. Interstitial fluid is a thin layer of fluid that surrounds the cells of the tissus below the skin \cite{djak2003, buck2008, benh2012}. That's the reason why there is a 5 to 10 minute delay in interstitial fluid glucose response to changes in blood glucose. This result of great importance will need to be compared to the Lyapunov time obtained in this study (see Sec. 4). Today, the three main manufacturers which propose devices with continuous glucose monitoring reading are Abbott (Freestyle Navigator II), Medtronic (MiniMed coupled with Veo-pump), and Novalab which offers the reader Dexcom G4 coupled to the insulin pump Animas Vibe. Whatever the system, it is composed of two parts. The glucose sensor which has a small, flexible tip that, inserted just under the skin, continuously measures the glucose concentration in the interstitial fluid, and stores data during several days. The blood glucose reader is used to scan the sensor and displays the current glucose reading based on the most recently updated glucose value during the latest hours of continuous glucose data. Some readers are also coupled to an external insulin pump (Paradigm VEOTM Medtronic or Animas VibeTM of Novalab). In this work, we have chosen to use the Abbott (Freestyle Navigator II) system because the blood glucose reader records the blood glucose variations continuously every minute all day long during fourteen consecutive days. This represents, for each patient, one value of glycaemia per minute during 24 hours and 14 days, i.e., 20,160 data. Thus, ten type 1 diabetes patients have accepted to provide us the recordings of their blood glucose during fourteen consecutive days so that they could be anonymously analyzed.

\subsection{Methods}

Following the works of Takens [1981], Sauer et al. [1991], and Abarbanel [1996], summarized in Kodba et al. [2005], we consider the reconstruction of the attractor in an $m$-dimensional phase space starting from the time series $\{x_1, \ldots, x_i, \ldots, x_T \}$ of blood glucose variations for each patient. Here $x_i$ denotes the glycaemia in $i^{th}$ minute. According to Takens [1981], the reconstructed attractor of the original system is given by the vector sequence

\begin{equation}
\label{eq1}
p\left( i \right) = \left( x_{t - (m - 1)\tau},\dots, x_{i-2\tau}, x_{i-\tau}, x_t \right)
\end{equation}

where $\tau$ and $m$ are the time delay and the embedding dimension, respectively. Takens' famous theorem states that for a large enough $m$, this procedure provides a one-to-one image of the original system. It follows that the attractor constructed according to equation (\ref{eq1}) will have the same dimension and Lyapunov exponents as the original system. To reconstruct the attractor successfully, pertinent values of $\tau$ and $m$ have to be accurately determined.

\subsubsection{Time Delay}

Two criteria are to be taken into account for the estimation of the time delay $\tau$:

\begin{itemize}

\item[-] $\tau$ has to be large enough because the information we get from measuring the value of $x$ at time $i + \tau$ should be significantly different from $x$ at time $i$.

\item[-] $\tau$ must not be larger than the time in which the system loses memory of its initial state. This is important for chaotic systems, which are unpredictable and lose memory of the initial state.

\end{itemize}

Fraser et al. [1986] defined the mutual information between $x_i$ and $x_{i + \tau}$ as a suitable quantity for determining $\tau$. The mutual information between $x_i$ and $x_{i + \tau}$ measures the quantity of information according to the following expression

\begin{equation}
\label{eq2}
I\left( \tau \right) = \sum\limits_{h} \sum\limits_{k} P_{h,k}\left( \tau \right) \log_2 \frac{P_{h,k}\left( \tau \right)}{P_{h}P_{k}}.
\end{equation}

Here $P_h$ and $P_k$ denote the estimated probabilities that supposes a value inside the $h^{th}$ bin and the $k^{th}$ bin, respectively and $P_{h,k}( \tau )$ is the estimated probability that $x_i$ is in $h^{th}$ bin and $x_{i + \tau}$ is in $k^{th}$ bin. The optimal choice for the time delay is given by the first minimum of $I(\tau)$, because $x_{i + \tau}$, then adds the largest quantity of information to the information at $x_i$.

\subsubsection{Embedding Dimension}

The method of false nearest neighbors proposed by Kennel et al. [1992] is used to determine the embedding dimension $m$. This method is based on the fact that points which are close in the reconstructed embedding space have to stay sufficiently close also during forward iteration. So the distance between two points of the reconstructed attractor cannot grow further as a threshold $R_{tr}$. Nevertheless, if an $i^{th}$ point has a close neighbor that does not fulfil this criterion, then this $i^{th}$ point is marked as having a false nearest neighbor. In order to calculate the false nearest neighbors, the following algorithm is used:

\begin{itemize}
\item[-] For a point $p(i)$ in the embedding space, we have to find a neighbor $p(j)$ for which $|| p(i) - p(j) || < \varepsilon$,  where $||\ldots||$ is the square norm and $\varepsilon$ is a small constant usually not larger than the standard deviation of data.

\item[-] The normalized distance $R_i$ between the points $p(i)$ and $p(j)$ is computed:

\begin{equation}
\label{eq3}
R_i = \frac{| x_{i + \tau} - x_{j + \tau} |}{|| p(i) - p(j) ||}
\end{equation}

\item[-] If $R_i$ is larger than the threshold $R_{tr}$, then $p(i)$ has a false nearest neighbor. According to Abarbanel et al. [1996], $R_{tr} = 15$ has proven to be a good choice for most data sets.

\end{itemize}

After having computed (with the algorithms quoted above) the optimal time delay $\tau$ and the embedding dimension $m$, the attractor can be successfully reconstructed.

\subsubsection{Correlation Dimension}

The correlation dimension is a generalization of the usual integer-valued dimension. The method used is called the method of correlation exponent. Suppose that we have data $x_1, x_2,\ldots$. Prepare the $m$-dimensional delay coordinates $p(t) = ( x_{t-(m-1)\tau},\ldots, x_{t-2\tau}, x_{t-\tau}, x_t)$. Let $N$ be the number of these delayed points $p(t)$. Define $n(r)$ as the number of pairs $(p(i), p(j))$, $i < j$, whose distance is smaller than $r$. Define then $C(r)$ to be the corresponding relative frequency:

\begin{equation}
\label{eq3}
C(r) = \dfrac{n(r)}{\binom{N}{2}}.
\end{equation}

Thus, $C(r)$, also called the correlation sum, represents the probability that a randomly chosen pair of points in the reconstructed phase space will be less than a distance $r$ apart. When $N$ approaches infinity, $C(r)$ is called the correlation integral. It is a measure of the spatial correlation between data points. It can be shown that when $N$ is large and $r$ small, $C(r)$ behaves approximately like a power function: $C(r) \approx \alpha r^\nu$, that is, $C(r)$ is proportional to $r^\nu$. Here, $\nu$ is called the correlation exponent. Because $\log C(r) \approx \log a + \nu \log r$, we see that $\log C(r)$ behaves linearly as a function of $\log r$. Furthermore, $\nu$ is the slope of $\log C(r)$ versus $\log r$.
To estimate the correlation dimension, we estimate $\nu$ for increasing $m$. If the correlation exponent saturates with increasing $m$, the system is considered to be chaotic and the saturation value is called the correlation dimension of the attractor. Thus, in estimating the correlation dimension, we look for an interval for $r$ where $\log C(r)$ versus $\log r$ behaves approximately linearly. Such an interval is called a scaling region. If the slopes of the curves in the scaling region are approximately constant for a range of values of $m$, this constant is an estimate of the correlation dimension. The saturation of the correlation exponent for increasing $m$ is a necessary condition for accepting an estimate as the correlation dimension \cite{Galka2000}. For chaotic data, the correlation dimension gives a fractional dimension for the strange attractor \cite{Grassberger1983}. Note that the presence of noise in the data makes the estimation of the correlation dimension much more uncertain. Noise may increase the estimate and may cause nonconvergence of the estimate for increasing $m$ \cite{Galka2000}. Define the noise level to be the ratio of the standard deviation of the noise to the standard deviation of the noise-free data. A small noise level of 2\% will typically make the estimation of the correlation dimension impossible! Thus, nonlinear noise reduction is important to noisy data. For a stochastic system (in contrast to a chaotic system), the correlation exponent increases without bound when the embedding dimension is increased. Thus, the correlation exponent can be used to distinguish between low-dimensional dynamics and randomness.
In the calculation of the correlation sum, we encounter one problem, namely dynamic correlation. To avoid the dynamic correlation, only pairs of points whose time distance is large enough are used. To this aim, Provenzale et al. [1992] have proposed the use of so-called space-time-separation plot. Once such time distance has been evaluated, the correlation sums are computed and so, the correlation dimension can be estimated.

\subsubsection{Determinism Test}

In order to confirm that type 1 diabetes is a chaotic phenomenon we have used the determinism test proposed by Kaplan et al. [1992]. The determinism test is of great important because it enables us to distinguish between deterministic chaos and irregular random behavior, which often resembles chaos. A  program by Perc [2006] enables to compute the determinism factor $\kappa$ which is substantially smaller than 1 for a system with a stochastic component while it is close to 1 for a system with a chaotic component. According to Kodba et al. [2005] ``the deterministic signature is still good enough to be preserved, so that the chaotic appearance of the reconstructed attractor cannot be attributed to stochastic influences.''

\subsubsection{Maximal Lyapunov Exponent}

The maximal Lyapunov exponent is a characteristic of the dynamical system and quantifies the strength of chaos. Indeed, in chaotic systems nearby trajectories diverge exponentially fast. The properly averaged exponent of this divergence is called the maximal Lyapunov exponent. It describes the average rate at which predictability of the system is lost. More than twenty years ago, Kantz [1994] presented the method of local divergence rates to calculate the maximal Lyapunov exponent (see also Kantz and Schreiber [2004]). This method estimates local divergence rates of nearby trajectories.
Let the data be $x_1, x_2,\ldots , x_T$. Choose a delay time $\tau$ and an embedding dimension $m \geqslant 2$ and prepare the $m$-dimensional delay coordinates $p(t) = ( x_{t-(m-1)\tau},\ldots, x_{t-2\tau}, x_{t-\tau}, x_t)$, $t = 1 + (m-1)\tau, \ldots, T$. Choose an $\epsilon > 0$, to be used later to find points of the data in an $\epsilon$-neighborhood. Choose an integer $\delta$ (typically, $\delta$ is between 0 and, say, 20). Choose a $t$ from the set $\mathcal{T}_\delta = \{ 1 + (m-1)\tau, \ldots, T-\delta \}$. Find all points $p(i)$ that are in the $\epsilon$-neighborhood of $p(t)$; let $\mathcal{U}_t$ be the set of these indices $i$. Calculate, for all $i \in \mathcal{U}_t$, the distance between $p(t)$ and $p(i)$ after the relative time $\delta$; this distance is defined as

\begin{equation}
\label{eq4}
dist\left(p(t), p(i); \delta  \right) = | x_{t + \delta} - x_{i + \delta} |.
\end{equation}

Thus, the distance is the absolute value of the difference between the scalars $x_{t + \delta}$ and $x_{i + \delta}$. Calculate then the logarithm of the mean of these distances. Do this for all $t \in \mathcal{T}_\delta$, and then calculate the mean of the calculated values, that is, the $\mathcal{T}_\delta$-mean of the logarithms of the $\mathcal{U}_t$-means of the distances. In this way, we get a value that we denote $S(\delta)$. A formula for $S(\delta)$ is as follows:

\begin{equation}
\label{eq5}
S\left( \delta \right) = \underset{t \in \mathcal{T}_\delta}{mean} \left(  \ln \left\{ \underset{i \in \mathcal{U}_t}{mean} \left[ dist \left( p(t), p(i); \delta \right) \right] \right\}  \right).
\end{equation}

It is a measure of the local divergence rate of nearby trajectories. Repeat these calculations for various values of $\delta$, such as $\delta = 0,\ldots, 20$. In this way we get the values of $S$ at discrete points. For chaotic dynamical systems, the $S(\delta)$ function behaves as follows:

\begin{itemize}

\item[-] For small values of $\delta$, $S(\delta)$ may behave a little irregularly.
\item[-] For intermediate values of $\delta$, $S(\delta)$ increases linearly. Such values of $\delta$ define a so-called scaling region. The slope of the linear growth is an estimate of the maximal Lyapunov exponent. A positive Lyapunov exponent is an indicator of chaotic behavior.
\item[-] For larger value of $\delta$, $S(\delta)$ tends towards a constant.

\end{itemize}

In the following section we have used the programs mentioned above for the determination of the time delay $\tau$, the embedding dimension $m$ and of course for the estimation of the correlation dimension, the maximal Lyapunov exponent as well as the Lyapunov time, {\it i.e.}, the inverse of the maximal Lyapunov exponent.

\section{Results}

We applied the method of reconstruction of the attractor in an $m$-dimensional space phase (recalled in the previous section) to the time series of blood glucose variations for each of our ten patients. In the first place, we considered the whole time series including days and nights. Nevertheless, because of the stochastic inputs induced by daily activities of any human being (work, eat, walk, sport,...), it has not been possible to discriminate chaos from noise. So, in the second place, we have decided to keep only the nights of these ten patients starting from 10:00 p.m. to 08:00 a.m. Therefore, for each patient, we have one value of glycaemia per minute during 10 hours, i.e., 600 data. But, numerical analysis methods imply the time series to have a sufficient length, typically comprising several thousand points or more. Consequently, we have concatenated the 14 nights of each patient in order to obtain for each of them nearly 8,400 data. Then, we applied the programs developed by Ruskeep\"{a}\"{a} and Perc to each of our ten patients. In the first subsection, we detail the procedure for patient 1. In subsection 3.2 we will only give the results for all the ten patients.

\subsection{Patient 1}

\subsubsection{Blood Glucose Data}

For the 14 concatenated nights of this first patient, starting from 10:00 p.m. to 08:00 a.m., we have $N = 8243$ data (see Fig. 1).

\subsubsection{Time Delay}

According to subsection 2.2.1, the optimal choice for the time delay is given by the first minimum of the average mutual information $I(\tau)$. For our data, there is no point of minimum (see Fig. 2). So, in this case, we can consider the value of $\tau$ at which the first substantial decrease of the average mutual information stops and a slower decrease begins. Thus, we can choose for the time delay say the value $\tau = 5$. We could also have chosen as the time delay another value like $10, 15$, or $20$. Consequently, we will use these four values of the time delay in the following to show that the results of our computations are invariant with respect to $\tau$.

\subsubsection{Embedding Dimension}

According to subsection 2.2.2, we use the method of false nearest neighbor to compute the embedding dimension. However, since we consider four possible values of the time delay, we have plotted in Fig. 3 the fraction of false nearest neighbors for $\tau = 5, 10, 15, 20$ as a function of the embedding dimension $m$ (the lowest curve is for $\tau = 5$ and the highest curve for $\tau = 20$). It appears that the fraction of false nearest neighbors does not drop to zero with any value of $m$ (whatever the value of $\tau$) but with $m = 5$ the fraction is almost zero.

These results are summarized in Table I. So, for patient 1 we can choose the embedding dimension $m=5$.

\subsubsection{Correlation Dimension}

According to subsection 2.2.3, we use the method of correlation exponent. First, to avoid the dynamic correlation we use space-time-separation plot proposed by Provenzale et al. [1992]. The plot on Fig. 4 indicates that if points are at least, say, 500 steps from each other, they will no more have dynamic correlation.

We calculate correlation sums by first using $\tau=5$, $m = 1,\ldots,12$, minimum time distant 500, and space distances $r=10^d$, where $d$ ranges from 0 to 2.5 so that we have a total of 25 values of $d$ (see Fig. 5).

It turns out that a scaling region is between the fifth and ninth value of $r$: there (between the two vertical lines, that is, for $r$ from somewhat below 3 to somewhat below 7) the $C(r)$-curves in Fig. 5 behave approximately linearly and, as Fig. 6 shows, the slopes of the $C(r)$-curves in the scaling region approach a saturation value, approximately 1.72. Indeed, for $m = 5, 6, 7$, and 8 the slopes are near to each other: 1.717, 1.720,1.723, and 1.725. The mean of these slopes is 1.721. This leads, for $\tau = 5$, to an estimate of 1.721 for the correlation dimension of the attractor.

These calculations for the correlation dimension was done by first using a Mathematica program to calculate the correlation sums (for a fixed value of $\tau$) and then applying a dynamic Mathematica interface to find a scaling region and the corresponding estimate of the correlation dimension. This dynamic interface contains sliders to manually adjust the start and end points of the possible scaling region and shows the slopes of the correlation sums, dynamically reflecting the current choice of the scaling region. So, to find a scaling region the sliders are adjusted until a promising region is found where the correlation sums behave approximately linearly and the slopes in the scaling region saturate with increasing $m$. If, in addition, the invariance with respect to $\tau$ can be
verified, this gives support to the estimated correlation dimension.

\subsubsection{Determinism Test}

While using Perc's program an estimation of the determinism factor is $\kappa = 0.84$.

\subsubsection{Maximal Lyapunov Exponent}

To estimate the maximal Lyapunov exponent, we use the method of local divergence rates. First, we choose the delay time $\tau=5$ and set $\epsilon=3$. The embedding dimension $m$ varies from 1 to 20. We obtain the values of $S(\delta)$ plotted in Fig. 7.

It turns out that a scaling region is for $\delta = 1,\dots,6$: there the $S(\delta)$-curves in Fig. 7 behave approximately
linearly (for larger values of $m$) and, as Fig. 8 shows, the slopes of the $S(\delta)$-curves in the scaling region are
approximately a constant for $m = 12,\ldots, 17$; the mean of these slopes is 0.0150. This leads, for $\tau = 5$, to an
estimate of 0.0150 for the maximal Lyapunov exponent.

The maximal Lyapunov exponent is invariant under the embedding procedure and also under different
values of $\epsilon$. Fig. 8 shows, for $\tau = 5$ and $\epsilon = 3$, the invariance with respect to the embedding dimension $m$ (the
slopes are approximately constant for $m = 12,\ldots, 17$). For $\tau = 5$, we also considered the values $\epsilon = 2.5$ and 2, and
we got the estimates 0.0141 and 0.0156. The estimates for the three values of $\epsilon$ are near to each other,
supporting, for $\tau = 5$, the invariance with respect to $\epsilon$. The mean of the three estimates is 0.0149.

We also estimated the maximal Lyapunov exponent for $\tau = 10, 15$, and 20. For each $\tau$, we considered the
values 3, 2.5, and 2 for $\epsilon$. For each $\tau$, the estimate was invariant with respect to $\epsilon$. The means of the estimates for
$\tau = 5, 10, 15$, and 20, were 0.0149, 0.0146, 0.0151, and 0.0145. These values are near to each other, supporting
the invariance with respect to $\tau$. The mean of the four estimates is 0.0147 (here, we used the original, more
accurate estimates). Thus, the final estimate of the maximal Lyapunov exponent is 0.0147. The Lyapunov time is
then $1/0.0147 \approx 68 minutes$.

These calculations for the maximal Lyapunov exponent done by first using a Mathematica program to calculate the values of $S(\delta)$ (for a fixed value of $\tau$ and $\epsilon$) and then applying a dynamic Mathematica interface to find a scaling region and the corresponding estimate of the maximal Lyapunov exponent.

In the next subsection, we summarize the main results for all the patients.

\subsection{Patient 1 to 10}

\subsubsection{Data Length}

The lengths of the data for the ten patients (see Table II) varied between 6892 and 8276 with a mean of 7961.4. For each patient, the data contains values of glucose for 14 concatenated nights. The lengths of the data vary somewhat because some values of the glucose have not been recorded by the glucose sensor. The lengths of the data turned out to be large enough for the numerical programs to work properly.

\subsubsection{Average Mutual Information: determination of the time delay $\tau$}

For the time delay $\tau$, the situation for all the ten patients was similar: the average mutual information did not have a local point of minimum. As examples of the time delay, we considered the values $\tau =5, 10, 15$, and 20 when we estimated the embedding dimension, the correlation dimension, and the maximal Lyapunov exponent.

\subsubsection{False Nearest Neighbors: determination of the embedding dimension $m$}

The embedding dimension $m$, estimated by the method of false nearest neighbors, varied between 5 and 20 for the four values of $\tau$. The fraction of false nearest neighbors did not drop exactly to zero but decreased to a small enough level (typically below 0.01 but above 0.001). The results are presented in Table III.

\subsubsection{Correlation Dimension}

The correlation dimension (see Table IV), estimated by the method of correlation exponent, varied between 1.19 and 2.61, the mean being 1.74. For each patient, the correlation dimension was estimated with the four values of the time delay $\tau$ (the values of the correlation dimension given in Table IV are the means of the four estimates got by using the four values of $\tau$).

\subsubsection{Determinism Test}

The estimation of the determinism factor $\kappa$ for the 14 concatenated nights of each of the 10 patients is presented in Table V.

Another program developed by BenSa\"{i}da [2015] has enabled to confirm the presence of a chaotic dynamics, as opposed to stochastic dynamics, in the 14 concatenated nights of each of our 10 patients.

\subsubsection{Maximal Lyapunov Exponent and Lyapunov Time}

The maximal Lyapunov exponent (see Table VI), estimated by the method of local divergence rates, varied between 0.0147 and 0.0353, the mean being 0.0215. For each patient, the exponent was estimated with the four values of $\tau$ and, for each $\tau$, the exponent was estimated with $\epsilon$ equal to 2, 2.5, and 3 (the values given in Table VI are the means of the resulting 12 estimates). The Lyapunov time (see Table VI) varied between 28 and 68 minutes, the mean being 52.5 minutes. The inverse of the final estimate of the Maximal Lyapunov exponent, 0.0215, is 46.5 minutes.

\section{Discussion}

Starting from a database often type 1 diabetes patients wearing a continuous glucose monitoring device, we have used numerical methods to analyze the continuous variations of the blood glucose every minute all day long during fourteen consecutive days. Nevertheless, because of the stochastic inputs induced by daily activities of any human being (work, eat, walk, sport, etc.), it has not been possible to discriminate chaos from noise in the whole time series including days and nights. So, we decided to keep only the fourteen nights of these ten recordings of the blood glucose, from 10:00 p.m. to 08:00 a.m. We then concatenated these recordings over fourteen nights and so we obtained, for each patient, nearly 8,400 data values; this amount of data was large enough for the computations.

We used the average mutual information to investigate the time delay $\tau$ for each of our ten patients. As the average mutual information did not have a local point of minimum, in later computations we considered, as examples, the values 5, 10, 15, and 20 for $\tau$. We were able to show that the estimates of the correlation dimension and the maximal Lyapunov exponent were invariant with respect to these values of $\tau$. The method of false nearest neighbors led to a determination of the embedding dimension $m$, a typical value being between, say, 5 and 12, the value varying with $\tau$ and with the patient in question.

The test of determinism provided a mean value 0.83 of the determinism factor for the ten patients. This value
is close to one and so supports the chaoticity of type 1 diabetes.

With the four values of $\tau$, we applied the method of correlation exponent to the ten time series. In each case, the correlation sum saturated with increasing $m$ and was then approximately constant in a scaling region, that is, for some values of $m$. This behavior supports the chaoticity of our data. The correlation dimension was, on the average, 1.74.

Finally, we estimated the maximal Lyapunov exponent with the method of local divergence rates. For each of the ten patients, we estimated the exponent for the four values 5, 10, 15, and 20 of $\tau$ and, for each $\tau$, for the three values 2, 2.5, and 3 of $\epsilon$. The estimates turned out to be invariant with respect to $\tau$ and $\epsilon$. For each patient, the mean of these twelve estimates is given in Table VI. The overall mean of these ten estimates is 0.0215. The corresponding estimate of the Lyapunov time is 46.5 minutes.

According to Feinberg and Floyd [1979], ``The notion that there exists in man a basic rest-activity cycle with a period of 90 min has been accepted by many investigators. This hypothesis was introduced by Kleitman (1963), who based it upon early observations (e.g. Dement \& Kleitman, 1957) that REM sleep in adults recurred at about 90-min intervals.'' More than ten years before this publication. Hartman [1968] proved that ``the mean cycle length for 15 normal adult subjects in our laboratory, studied for eight or more nights each, was 95.8 minutes, with a standard deviation of 8.7 minutes.'' Thus, the ``sleep-dream'', also called rapid eye movement (REM)-sleep, recurred in adults at about 90 minutes intervals and is linked to real time, i.e., to the circadian cycle.
Since we have found an estimate of the Lypaunov time equal to 46.5 minutes, which corresponds nearly to half the sleep cycle or rest-activity cycle, it seems to indicate that some correlations could be highlighted between blood glucose variations in type 1 diabetes patients and their REM-sleep. This would be subject of a new protocol of research in which type 1 diabetes patients wearing a continuous glucose monitoring device will undergo a polysomnography.

\begin{acknowledgments}
Matja{\v z} Perc was supported by the Slovenian Research Agency (Grants J1-7009 and P5-0027).
\end{acknowledgments}

\newpage
\section{Figures}

\begin{figure}[H]
\centerline{\includegraphics[width = 13cm]{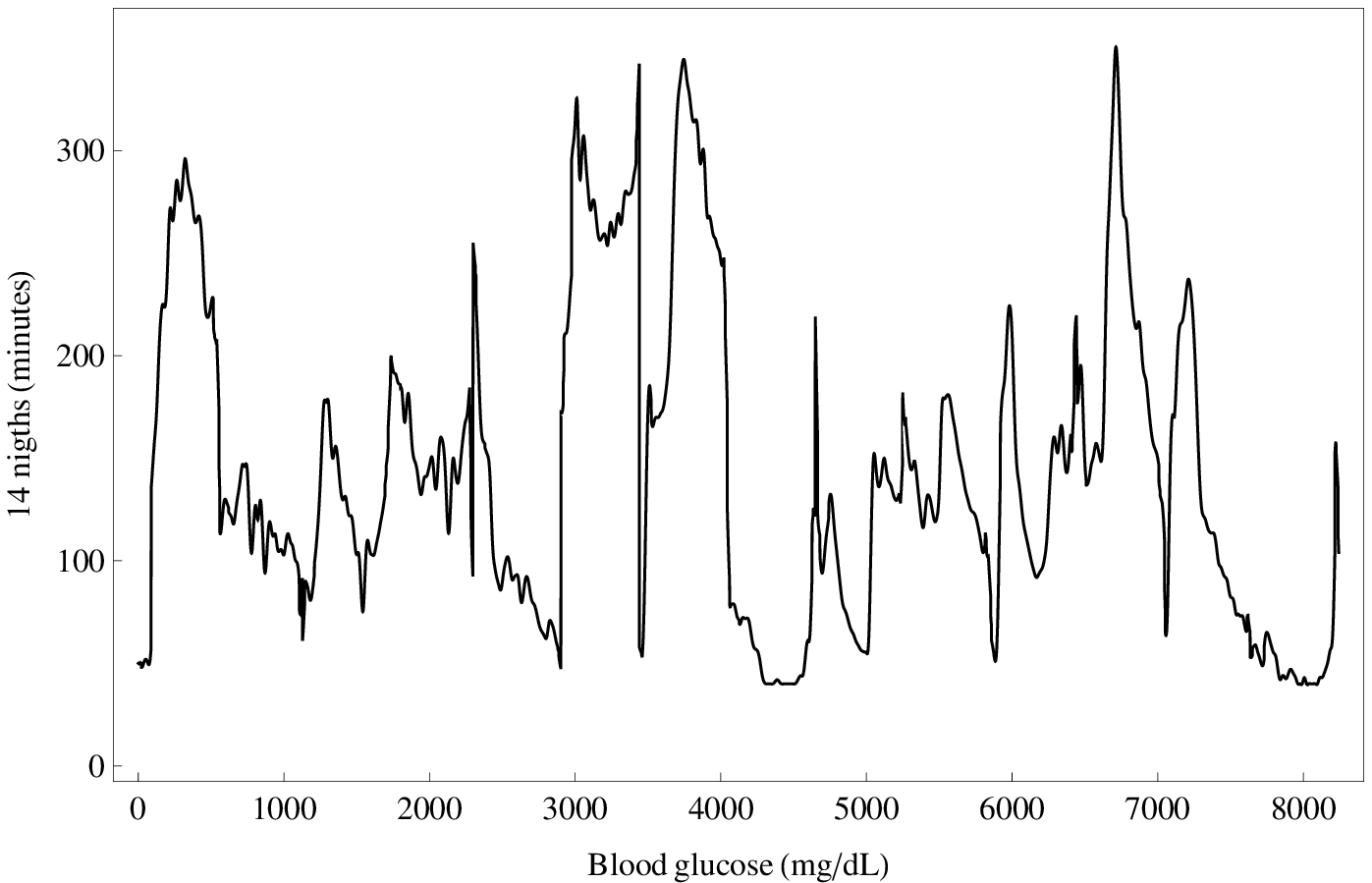}}
\caption{Blood glucose variations during the 14 concatenated nights of patient 1.}
\label{fig1}
\end{figure}

\begin{figure}[H]
\centerline{\includegraphics[width = 13cm]{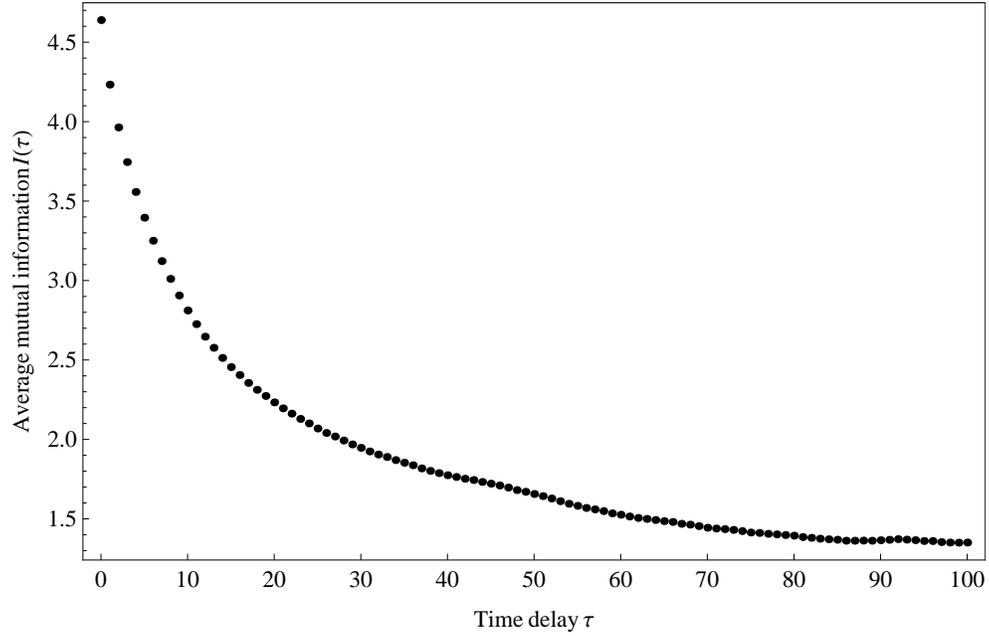}}
\caption{Average mutual information $I(\tau)$ of patient 1.}
\label{fig2}
\end{figure}

\begin{figure}[H]
\centerline{\includegraphics[width = 13cm]{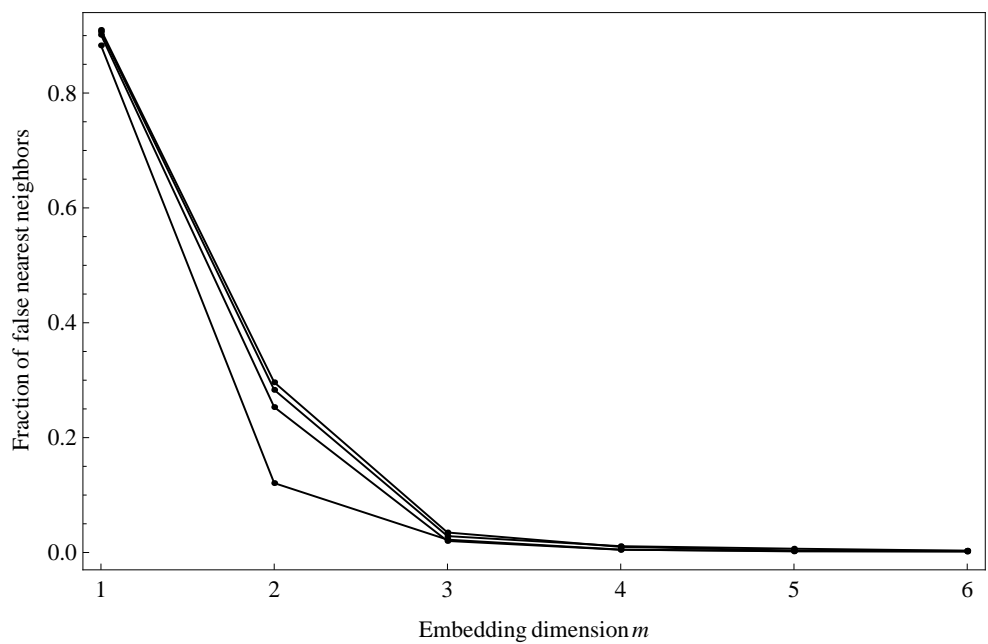}}
\caption{Fraction of false nearest neighbors as function of the embedding dimension $m$ of patient 1.}
\label{fig3}
\end{figure}

\begin{figure}[H]
\centerline{\includegraphics[width = 16cm]{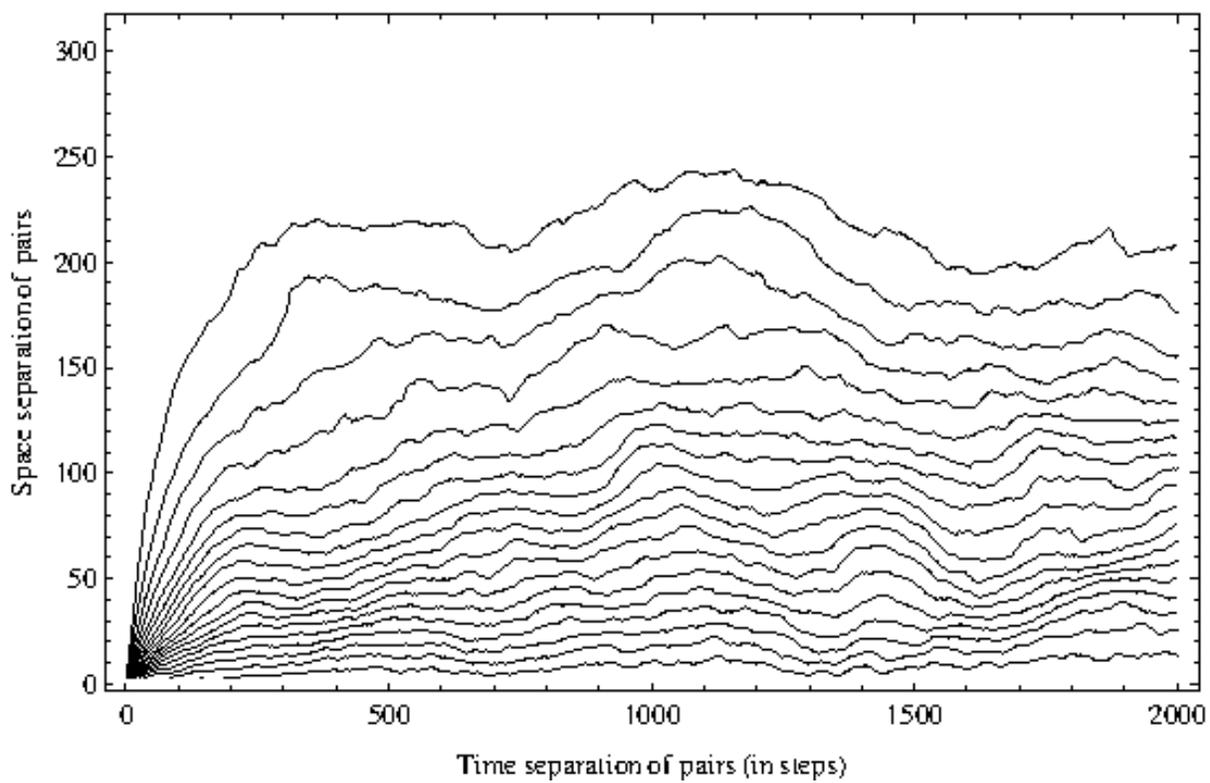}}
\caption{Space-time-separation plot of patient 1.}
\label{fig4}
\end{figure}

\begin{figure}[H]
\centerline{\includegraphics[width=12cm]{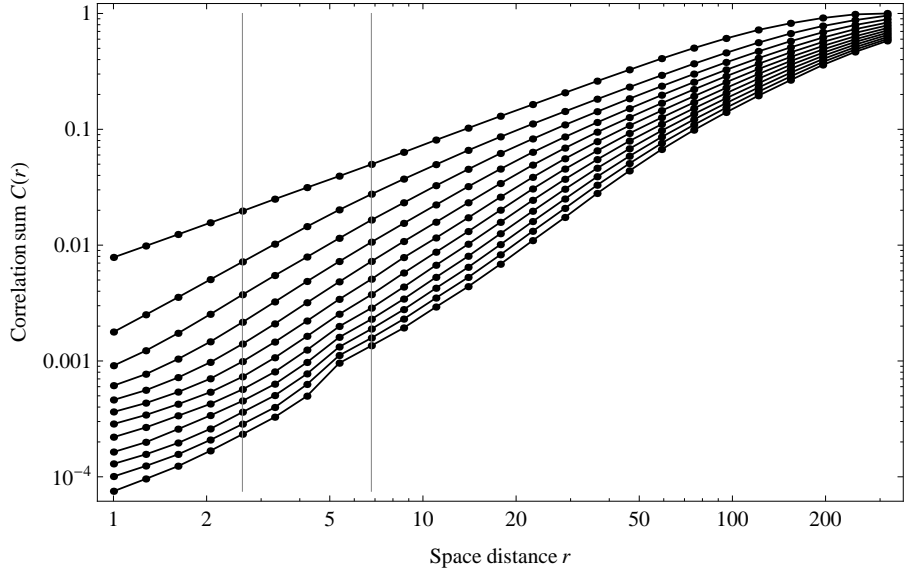}}
\caption{Values of correlation sums $C(r)$ of patient 1.}
\label{fig5}
\end{figure}

\begin{figure}[H]
\centerline{\includegraphics[width=12cm]{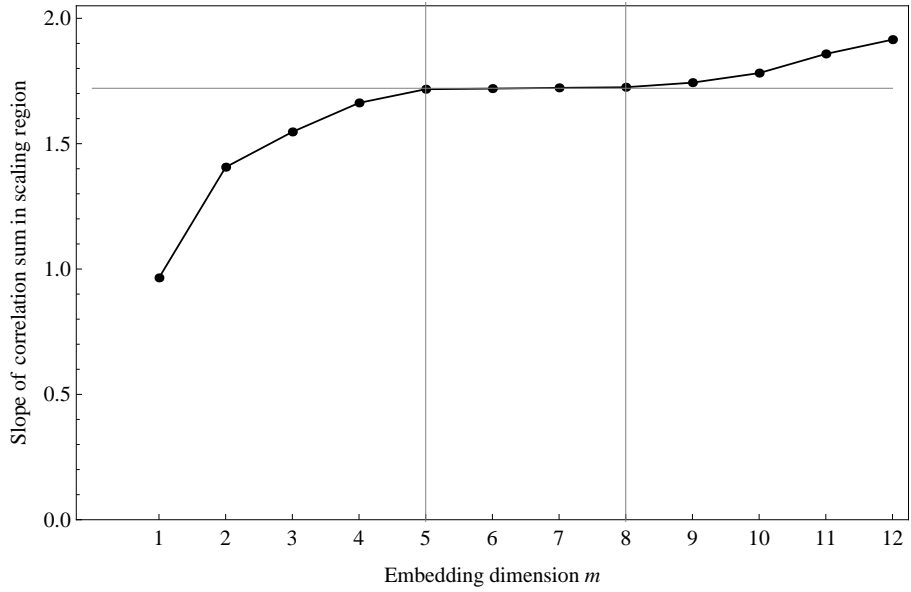}}
\caption{Slope of linear fits of the correlation sums in the scaling region for patient 1.}
\label{fig6}
\end{figure}

\begin{figure}[H]
\centerline{\includegraphics[width=13cm]{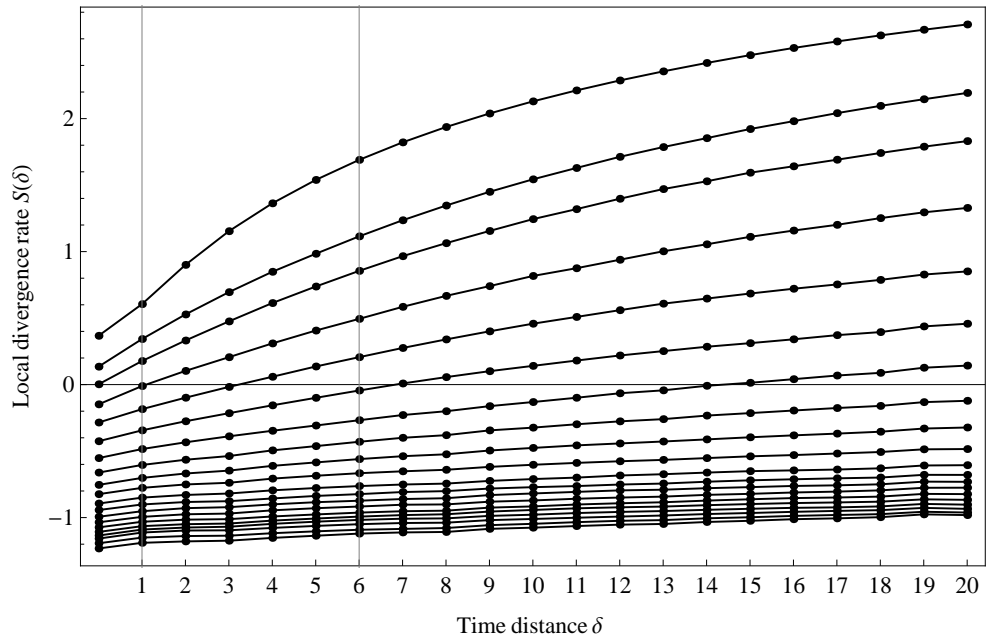}}
\caption{Values of $S(\delta)$ of patient 1.}
\label{fig7}
\end{figure}

\begin{figure}[H]
\centerline{\includegraphics[width=12cm]{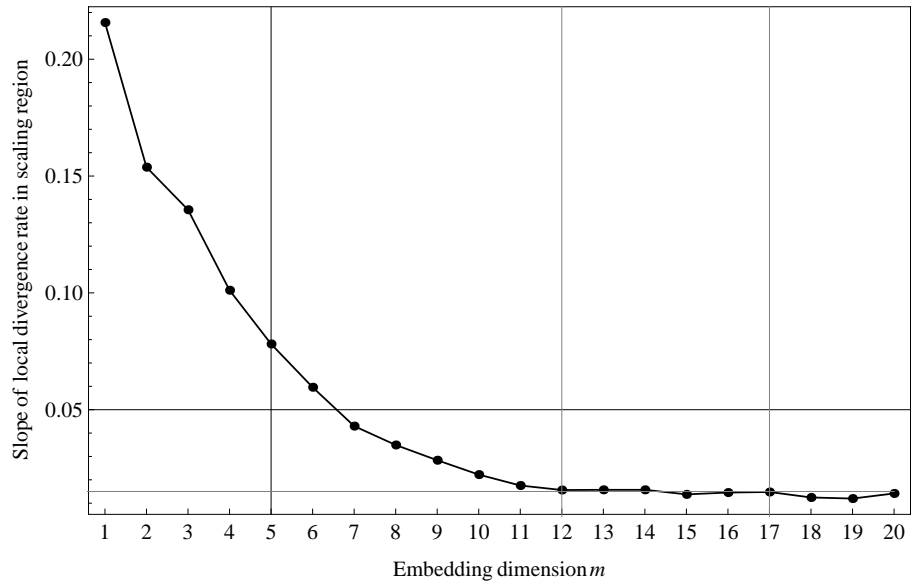}}
\caption{Slopes of linear fits of the $S(\delta)$-values in the scaling region for patient 1.}
\label{fig8}
\end{figure}

\newpage
\section{Tables}

\begin{table}[H]
\centering
\caption{Fraction of false nearest neighbors as a function of the embedding dimension $m$}.
{\begin{tabular}{c c c}\\[-2pt]
\toprule
Fraction of \textit{fnn} & $m$ & $\tau$ \\[6pt]
\hline\\[-2pt]
\hspace{2cm} 0.0019 \% \hspace{2cm} & \hspace{2cm} 5 \hspace{2cm} & \hspace{2cm} 5 \hspace{2cm} \\[1pt]
0.0027 \% & 5 & 10 \\[2pt]
0.0066 \% & 5 & 15 \\[2pt]
0.0034 \% & 5 & 20\\[1pt]
\botrule
\end{tabular}}
\end{table}

\begin{table}[H]
\centering
\caption{Data length for the 14 concatenated nights of each patient.}
{\begin{tabular}{c c c c c c c c c c c c}\\[-2pt]
\toprule
Patient & 1 & 2 & 3 & 4 & 5 & 6 & 7 & 8 & 9 & 10 & $\langle N \rangle$ \\[6pt]
\hline\\[-2pt]
N & 8243 & 8141 & 8226 & 6892 & 7387 & 8276 & 7843 & 8161 & 8199 & 8246 & 7961.4 $\pm$ 464.9 \\[1pt]
\botrule
\end{tabular}}
\end{table}

\begin{table}[H]
\centering
\caption{Fraction of false nearest neighbors as a function of the embedding dimension $m$.}
{\begin{tabular}{c c c c}\\[-2pt]
\toprule
\hspace{1cm} Patient \hspace{1cm} & \hspace{1cm} $\tau$ \hspace{1cm} & \hspace{1cm} \% \textit{fnn} \hspace{1cm} & \hspace{1cm} $m$ \hspace{1cm} \\[6pt]
\hline\\[-2pt]
1 & \{5, 10, 15, 20\} & \{0.0019 \%, 0.0027 \%, 0.0066 \%, 0.0034 \%\} & \{5,5,5,5\}\\[1pt]
2 & \{5, 10, 15, 20\} & \{0.0085 \%, 0.0110 \%, 0.0110 \%, 0.0120 \%\} & \{5,9,9,9\}\\[1pt]
3 & \{5, 10, 15, 20\} & \{0.0027 \%, 0.0027 \%, 0.0046 \%, 0.0021 \%\} & \{5,5,9,9\}\\[1pt]
4 & \{5, 10, 15, 20\} & \{0.0032 \%, 0.0031 \%, 0.0110 \%, 0.0094 \%\} & \{6,6,6,20\}\\[1pt]
5 & \{5, 10, 15, 20\} & \{0.0041 \%, 0.0057 \%, 0.0064 \%, 0.0049 \%\} & \{5,6,16,12\}\\[1pt]
6 & \{5, 10, 15, 20\} & \{0.0061 \%, 0.0072 \%, 0.0074 \%, 0.0075 \%\} & \{5,8,14,13\}\\[1pt]
7 & \{5, 10, 15, 20\} & \{0.0029 \%, 0.0032 \%, 0.0033 \%, 0.0034 \%\} & \{6,6,17,11\}\\[1pt]
8 & \{5, 10, 15, 20\} & \{0.0026 \%, 0.0022 \%, 0.0026 \%, 0.0024 \%\} & \{6,5,5,7\}\\[1pt]
9 & \{5, 10, 15, 20\} & \{0.0024 \%, 0.0043 \%, 0.0052 \%, 0.0043 \%\} & \{6,6,10,10\}\\[1pt]
10 & \{5, 10, 15, 20\} & \{0.0022 \%, 0.0027 \%, 0.0024 \%, 0.0019 \%\} & \{8,7,13,11\}\\[1pt]
\botrule
\end{tabular}}
\end{table}

\begin{table}[H]
\centering
\caption{Estimates of the correlation dimension for each patient.}
{\begin{tabular}{c c c c c}\\[-2pt]
\toprule
Patient & $\tau$ & Values of $r$ in the scaling region & Correlation Dimension $\nu$ & $\langle \nu \rangle$ \\[6pt]
\hline\\[-2pt]
1 & \{5, 10, 15, 20\} & \{[5,9],[6,10],[2,9],[1,14]\} & \{1.72,1.71,1.71,1.74\} & 1.72 \\[1pt]
2 & \{5, 10, 15, 20\} & \{[16,17],[16,17],[15,16],[13,15]\} & \{1.19,1.22,1.19,1.17\} & 1.19 \\[1pt]
3 & \{5, 10, 15, 20\} & \{[10,12],[9,14],[4,7],[6,10]\} & \{2.42,2.66,2.69,2.65\} & 2.61 \\[1pt]
4 & \{5, 10, 15, 20\} & \{[16,20],[13,18],[13,17],[12,18]\} & \{1.32,1.34,1.40,1.43\} & 1.37 \\[1pt]
5 & \{5, 10, 15, 20\} & \{[18,19],[13,15],[13,18],[12,17]\} & \{1.36,1.39,1.36,1.37\} & 1.37 \\[1pt]
6 & \{5, 10, 15, 20\} & \{[13,16],[11,12],[10,12],[9,10]\} & \{1.38,1.47,1.48,1.47\} & 1.45 \\[1pt]
7 & \{5, 10, 15, 20\} & \{[15,17],[12,15],[11,14],[5,8]\} & \{1.89,1.91,1.90,1.89\} & 1.45 \\[1pt]
8 & \{5, 10, 15, 20\} & \{[8,10],[5,6],[7,8],[2,7]\} & \{2.19,2.16,2.23,2.18\} & 2.19 \\[1pt]
9 & \{5, 10, 15, 20\} & \{[12,22],[10,19],[10,12],[8,10]\} & \{1.78,1.79,1.79,1.81\} & 1.79 \\[1pt]
10 & \{5, 10, 15, 20\} & \{[12,14],[9,12],[4,8],[4,7]\} & \{2.26,2.25,2.27,2.30\} & 2.27 \\[1pt]
\botrule
\end{tabular}}
\end{table}

\begin{table}[H]
\centering
\caption{Determinism factor $\kappa$ for the 14 concatenated nights of each patient.}
{\begin{tabular}{c c c c c c c c c c c c}\\[-2pt]
\toprule
Patient & 1 & 2 & 3 & 4 & 5 & 6 & 7 & 8 & 9 & 10 & $\langle \kappa \rangle$ \\[6pt]
\hline\\[-2pt]
$\kappa$ & 0.84 & 0.84 & 0.82 & 0.82 & 0.85 & 0.82 & 0.83 & 0.79 & 0.85 & 0.82 & 0.82 $\pm$ 0.018 \\[1pt]
\botrule
\end{tabular}}
\end{table}

\begin{table}[H]
\centering
\caption{Estimates of the $\langle MLE \rangle$ and of the $\langle LT \rangle$ for each patient.}
{\begin{tabular}{c c c c c c}\\[-2pt]
\toprule
Patient & $\tau$ & $\epsilon$ & MLE & $\langle MLE \rangle$ & $\langle LT \rangle$ \\[6pt]
\hline\\[-2pt]
1 & \{5, 10, 15, 20\} & \{3,2,2.5\} & \{0.0149,0.0146,0.0151,0.0145\} & 0.0147 & 68' \\[1pt]
2 & \{5, 10, 15, 20\} & \{3,2,2.5\} & \{0.0338,0.0340,0.0338,0.0327\} & 0.0336 & 30' \\[1pt]
3 & \{5, 10, 15, 20\} & \{3,2,2.5\} & \{0.0150,0.0146,0.0149,0.0148\} & 0.0148 & 68' \\[1pt]
4 & \{5, 10, 15, 20\} & \{3,2,2.5\} & \{0.0151,0.0154,0.0142,0.0146\} & 0.0148 & 68' \\[1pt]
5 & \{5, 10, 15, 20\} & \{3,2,2.5\} & \{0.0326,0.0323,0.0327,0.0320\} & 0.0324 & 31' \\[1pt]
6 & \{5, 10, 15, 20\} & \{3,2,2.5\} & \{0.0352,0.0356,0.0352,0.0348\} & 0.0353 & 28' \\[1pt]
7 & \{5, 10, 15, 20\} & \{3,2,2.5\} & \{0.0202,0.0203,0.0189,0.0196\} & 0.0198 & 51' \\[1pt]
8 & \{5, 10, 15, 20\} & \{3,2,2.5\} & \{0.0177,0.0174,0.0180,0.0175\} & 0.0177 & 57' \\[1pt]
9 & \{5, 10, 15, 20\} & \{3,2,2.5\} & \{0.0161,0.0160,0.0162,0.0160\} & 0.0161 & 62' \\[1pt]
10 & \{5, 10, 15, 20\} & \{3,2,2.5\} & \{0.0163,0.0164,0.0161,0.0164\} & 0.0162 & 62' \\[1pt]
\botrule
\end{tabular}}
\end{table}

\end{document}